\documentclass[twoside,reqno]{mbook}
\usepackage{epsfig,cite}
\usepackage{amssymb,amsmath}
\usepackage{times}
\begin{document}

\sloppy \raggedbottom

\setcounter{page}{1}
\newpage
\setcounter{figure}{0}
\setcounter{equation}{0}
\setcounter{footnote}{0}
\setcounter{table}{0}
\setcounter{section}{0}

\title{Description of the Nuclear Octupole and Quadrupole Deformation}

\runningheads{Description of the Nuclear Octupole and Quadrupole
Deformation}{P.G.~Bizzeti, A.M.~Bizzeti--Sona}

\begin{start}
\author{P.G. Bizzeti}{1,2}, \coauthor{A.M. Bizzeti--Sona}{1,2}

\address{Dipartimento di Fisica, Universit\`a di Firenze, Italy}{1}

\address{Istituto Nazionale di Fisica Nucleare, Sezione di Firenze}{2}

\begin{Abstract}
A parametrization of octupole plus quadrupole deformation, in terms
of intrinsic variables defined in the rest frame of the overall 
tensor of inertia, is presented and discussed. The model
is valid for situations close to the axial symmetry, but non
axial deformation parameters are not frozen to zero. The properties of
the octupole excitations in the deformed Thorium isotopes $^{226}$Th,
$^{228}$Th are interpreted in the frame of this model. 
A tentative interpretation of octupole oscillations in nuclei close
to the X(5) symmetry, in terms of an exactly separable potential, is
also discussed.

\end{Abstract}
\end{start}

\section[]{Introduction}
\label{S:1}
We present here a formalism to describe the simultaneous
octupole and quadrupole deformations of the nuclear surface,
close to but not coincident with the axial symmetry limit,
in the frame of Bohr hydrodynamical model. This scheme has
been developed in order to discuss phase--transition
phenomena in the octupole degree of freedom, a subject which
-- as it will be explained in the following Section~\ref{S:2} --
appeared as a natural development of the experimental and
theoretical researches of the group of Florence in the recent
years.

Sections ~\ref{S:3} and ~\ref{S:4} describe the basis of 
this model and its application to the phase transition between
octupole oscillations around a permanently deformed 
reflection--symmetric shape and rigid rotation of a 
reflection--asymmetric rotor. These parts summarize 
the results contained in a recent
paper in the Physical Review C~\cite{biz0}.
Finally, some new results  are reported in the Section~\ref{S:5},
while in Section~\ref{S:6} a simple model, based on a separable 
potential, is discussed and used to tentatively interpret the octupole 
oscillations in nuclei close to the X(5) critical point.

\section[]{Some history}
\label{S:2}
Our interest on the octupole excitations dates at least from the
late '90s, when we have searched and found evidence for
two--octupole--phonon excitations in $^{144}$Nd and 
$^{146}$Sm~\cite{biz146}
and we have long searched, but not found, three--octupole--phonon 
excitations in $^{148}$Gd~\cite{biz148}.

Later, we have been impressed by Iachello results~\cite{iac1,iac2}
on the symmetries
at the critical point, such as E(5) and X(5) (other symmetries,
like Y(5) and Z(5), have been proposed later~\cite{iac3,z0}).
The first example of X(5) symmetry was identified in 
$^{152}$Sm~\cite{casten},
a well known transitional nucleus. We had some experience of another
transitional region, that of Mo and Tc isotopes: so, we could
identify another possible X(5) candidate~\cite{biz104} in the $^{104}$Mo
isotope~\footnote{Level energies and $\gamma$ branching ratios
in $^{104}$Mo are in
excellent agreement with the X(5) model. 
Later measurements~\cite{hutter} of mean lives
showed, however, that $B$(E2) values are at variance 
with the model predictions.}.
We also observed that the X(5) model was able to account for not
only the ground--state band, but also the excited $\gamma$ bands,
both in  $^{104}$Mo and in $^{152}$Sm. This fact is not trivial,
because the negative parity bands, present at low excitation in 
$^{152}$Sm, show a very different behaviour. We shall come back
to this point in the following.

We also noted some regularities in the proton and neutron numbers
of these two X(5) nuclei. In fact,  $^{104}$Mo has $Z=42,\ N=62$,
 $^{152}$Sm has $Z=62,\ N=90$: one could suspect that another phase
transition takes place also for 
$Z\approx 90$.

Actually, the standard indicators of quadrupole collectivity, 
and in particular the ratio $R_2=E(4^+)/E(2^+)$ give indication of 
a phase transition in the Ra ($Z=88$) and Th ($Z=90$) isotopic
chains. Heavier isotopes have rotational character, while the lighter
ones appear to be vibrational (or non collective). Moreover,
the energies of positive parity levels in the ground--state band of 
$^{224}$Th
and $^{224}$Ra show an impressive agreement with the predictions of 
the X(5) model~\cite{biz1,biz2}. 
However, in addition to the positive parity band,
these nuclei possess a odd-$J$, negative parity band which starts with
a $1^-$ level lying slightly above the $2^+$ and merges with the
positive--parity one at $J\approx 5$ or 6. This means that octupole
degrees of freedom are important in these nuclei, and their effect
must be considered when the evolution of the nuclear shape is followed
along the isotopic chain.
We observe that $^{230}$Th and heavier isotopes of Th give evidence of
octupole vibrations combined with the rotation of a deformed (but
reflection symmetric) core: rotational--like bands are built over
the one--octupole--phonon states with different values of $K$ (the 
angular--momentum component along the approximate symmetry axis).
In lighter isotopes, the excitation energy of the $1^-$ band head 
of the $K^\pi=0^-$ band 
decreases well below those of the other octupole bands, and 
approaches the rotational--energy value. Eventually, one 
observes an alternate--parity band which approaches (but never
reaches) the behaviour of a rigid, rotation asymmetric rotor,
typical of binary asymmetric molecules. In conclusion, a phase 
transition for the octupole degrees of freedom seems to take 
place in the Th and Ra isotopes, not far from the critical point
of the quadrupole one, but in the opposite direction: with octupole
vibration where the quadrupole deformation is stable, and vice-versa.
We can also observe that only the $K^\pi=0^-$ band seems to change
substantially along the isotopic chain. This means that the relevant
quadrupole and octupole degrees of freedom are those related to
axially symmetric, or quasi--axially--symmetric, deformations.

At this point, one needs a theoretical scheme able to describe 
the evolution of quadrupole and octupole deformations, close to the 
axial symmetry, across the critical point of the phase transition.

In the frame of the algebraic approach, new developments of the
Interacting Vector Boson Model have been reported at this 
workshop~\cite{ivbm}. The $spdf$ extension of the
IBM~\cite{spdf} is also able to account for 
negative--parity
excitations. This model, however, is not the most suitable to describe
the phase transitions.

In the frame of the geometrical approach, the most complete treatment
is the one proposed by Donner and Greiner~\cite{donner}. They use the
Bohr intrinsic frame, defined by the symmetry axes of the quadrupole,
and refer to this frame the seven amplitudes of the octupole mode
(the overall tensor of inertia being {\em not} diagonal in this
reference frame).
This approach, however, is useful only when all the octupole amplitudes 
are small compared to the quadrupole deformation.
Other models~\cite{models,jolos,modelsr} frozen to zero part of the 
dynamical  variables, and are
usually limited to the axially symmetric case: what is probably
enough in the presence of a stable quadrupole deformation, but not
necessarily when the quadrupole deformation approaches zero.

Finally, we have noted a very interesting work by Wexler
and Dussel~\cite{wexler}, which shows that it is possible to define an
intrinsic reference frame for the octupole mode, in which
the tensor of inertia is diagonal. In this
frame, the seven octupole amplitudes can be parametrized in terms of
four intrinsic variables (as the five quadrupole amplitudes
are parametrized in term of $\beta_2$ and $\gamma_2$ in the
Bohr representation). This approach, however, is valid for the
octupole mode alone.

In conclusion: in order to describe quadrupole and octupole
deformation,
close to but not coincident with the axial symmetry and with the
octupole amplitude not necessarily small compared with the quadrupole
one, we were
obliged to introduce our own parametrization scheme.

\nopagebreak
\nopagebreak
\section[]{The basis of the model}
\label{S:3}

We must start, as usual, from the definition of the nuclear surface in
polar coordinates, in the laboratory frame:
\begin{equation}
r(\theta ,\phi ) = R_0 \Big[ 1 +\!\sum _{\lambda =2,3}\phantom{mj} 
\hspace*{-7mm}\sum _{\phantom{m}\mu = -\lambda, \lambda}\hspace{-2mm} 
\alpha ^{(\lambda )}_\mu Y^*_{\lambda, \mu}(\theta ,\phi ) 
\Big] ,\hspace{7mm} 
\alpha ^{(\lambda )}_{-\mu}= (-)^\mu {\alpha _\mu^{(\lambda )}}^*
\label{E:3.1}
\end{equation}
We limit our model space to quadrupole and octupole, and we assume
that the deformation is small enough not to need to introduce a
monopole or dipole term to keep constant the nuclear volume and the 
center--of--mass position.
For the expression of the kinetic energy, we use that of the Bohr
hydrodynamical model,
$T=\frac{1}{2}\sum_\lambda B_\lambda \sum_\mu 
|\dot{\alpha}^{(\lambda )}_\mu |^2\ .$
Then, the laboratory amplitudes $\alpha ^{(\lambda )}_\mu$ are
expressed
in terms of the corresponding amplitudes in a proper intrinsic frame
and of the Euler angles $\theta_1,\  \theta_2,\ \theta_3$ 
defining the orientation of the intrinsic frame in the lab,
\begin{equation}
\sqrt{B_\lambda}\ \alpha^{(\lambda)}_\mu =
\sum_\nu a^{(\lambda)}_\nu {D^{(\lambda )}_{\mu \nu}}^*(\theta_i )
\label{E:3.2}
\end{equation}
\vspace{-3mm}

\noindent
where the $D^{(\lambda )}$ are Wigner matrices. Note that, in
order
to simplify the notations, the inertia parameter $B_\lambda$ has been
included in our definition of $a^{(\lambda)}$.
In terms of the new variables, the expression of the kinetic energy
splits in three parts: a {\em vibrational} term 
$ T_{\rm vib}=\frac{1}{2}
\sum _{\lambda ,\mu} | \dot{a}^{(\lambda )}_\mu | ^2$;
a  {\em rotational} term $ T_{\rm rot}=\frac{1}{2} \sum_{k , k^\prime} 
{\mathcal{ J}_{k k^\prime}}q_k q_{k^\prime}$,  
quadratic in the intrinsic components $q_k$ of the angular
velocity;
and a coupling term $ T_{\rm coup}$, 
 not present in the Bohr model for
quadrupole motion:  
$T_{\rm coup}=i
\sqrt{21}
\ [ q^{(1)} \otimes 
[a^{(3)} \otimes \dot{a}^{(3)} 
] ^{(1)} 
] ^{(0)}_0$.
For the non diagonal components of the tensor of inertia,
the hydrodynamical model gives
\vspace{-4mm}

\noindent
\begin{eqnarray}
{\mathcal{ J}_{13} }\!&{+}&\!{
i\ \mathcal{ J}_{23}}\ =\ \sum_\lambda  {C_2(\lambda)}
\ \left[  a^{(\lambda )} \otimes  a^{(\lambda )}
\right]^{(2)}_1\cr 
{\mathcal{ J}_{23}}\!&=&
\!\sum_\lambda  {C_2(\lambda)}
\ {\rm Im} \left[  a^{(\lambda )} \otimes  a^{(\lambda )}
\right]^{(2)}_2 
\label{E:33}
\end{eqnarray}
\vspace{-3mm}

\noindent
where $C_2(2)=-\sqrt{21}$, $C_2(3)=\sqrt{126}$.
It is relatively easy to find a parametrization which automatically
sets to zero the non-diagonal products of inertia, as far as quadrupole
and octupole are treated separately. For the quadrupole, it is the
classical one by Bohr, 
$a^{(2)}_0 = \beta_2 \cos \gamma_2$, 
$a^{(2)}_1 = 0$ 
$a^{(2)}_2 = \sqrt{1/2}\ \beta_2\ \sin \gamma_2$. 
For the octupole, we assume a parametrization similar to that of
Wexler and Dussel~\cite{wexler}:
\begin{eqnarray}
a^{(3)}_0 &=& \beta_3 \ \cos \gamma_3 \cr
a^{(3)}_1 &=& -  (5/2)\ \left( X + i Y \right)\ \sin \gamma_3\\
a^{(3)}_2 &=& \sqrt{1/2}\ \beta_3\ \sin \gamma _3  \cr 
a^{(3)}_3 &=& X \left[ \cos \gamma_3 + (\sqrt{15}/2)\ \sin \gamma_3
\right] \nonumber 
+ i\ Y \left[ \cos \gamma_3 - (\sqrt{15}/2)\ \sin \gamma_3
\right] \ .
\label{E:3.4}
\end{eqnarray}
In both cases, the $a^{(\lambda)}_\mu$ amplitudes with $|\mu |=2$ are 
real, while the $|\mu
|=1$ terms are either zero or small of the second order, if we
consider small of the first order the other non-axial amplitudes.

The problem arises when both quadrupole and octupole terms are
present,
since the principal axes of the quadrupole and of the octupole tensor
of inertia do not necessarily coincide.

We want to use, as in the Bohr model, an intrinsic frame defined by
the principal axes of the tensor of inertia. We must therefore impose
the conditions ${\cal J}_{ij} =0$ for $i\neq j$. This is a set of 3 
non linear equations, but we can linearize them if we assume that
non axial amplitudes are small in comparison with the axial ones.
One obtains, up to the first order in the
small amplitudes,
\begin{eqnarray}
\mathcal{J}_{12}&=& - 2\sqrt{6} \left( \beta_2 {\rm Im}\ a^{(2)}_2+
\sqrt{5}\ \beta_3 {\rm Im}\ a^{(3)}_2 \right) = 0\cr 
\mathcal{J}_{13}&+&i \mathcal{J}_{23}\ = 
\ \sqrt{6}\ \left( \beta_2 a^{(2)}_1+
\sqrt{2}\ \beta_3 a^{(3)}_1 \right) = 0
\label{E:3.5}
\end{eqnarray}
 which are automatically verified if we put
\begin{equation}
\begin{array}[c]{lp{0.1mm}lp{0mm}l}
a^{(2)}_1=-c_1\sqrt{2}\ \beta_3\left(\eta_c + i \zeta_c \right) && 
a^{(3)}_1=c_1 \beta_2 \left(\eta_c + i\zeta_c \right) \\[8pt] 
{\rm Im}\;a^{(2)}_2= -c_2 \sqrt{5}\ \xi_c &&
{\rm Im}\;a^{(3)}_2=c_2 \beta_2\ \xi_c 
\end{array}
\label{E:3.6}
\end{equation}
with the new parameters $\eta_c,\ \zeta_c$ and $\xi_c$ small of the 
first order. The factors $c_1$, $c_2$ are -- at the moment -- arbitrary
functions of $\beta_2$ and $\beta_3$. 

Now, we can introduce the matrix of inertia $\cal G$, which relates the
classical kinetic energy to the components $u_i$ of the generalized
velocity vector:
\begin{equation}T= \frac{1}{2} \sum {\cal G}_{ik} u_i u_k
\label{E:3.7}
\end{equation}
where, in our case, $u=\{\dot{\beta}_2,\ \dot{\gamma}_2,\
\dot{\beta}_3,
\ \dot{\gamma}_3,\ \dot{X},\ \dot{Y},\ \dot{\xi},
\ \dot{\eta},\ \dot{\zeta},\ q_1,\ q_2,\ q_3\}$. 
The matrix  $\cal G$ and its determinant $G$ play an important role in
the quantization of the kinetic energy, according to the
Pauli procedure~\cite{pauli}. 

If we want that the results of the Bohr model be obtained at the limit
of $\beta_3 \ll \beta_2$, we must verify that this happens, first of
all, for the determinant $G$. 
Now, the simplest possible choice of constant factors $c_1$ ad $c_2$ in
the Eq.~\ref{E:3.6} would give $G\propto \beta_2^{14}$, while in the Bohr
model it should be $\propto \beta_2^8$. A better choice would be
\begin{equation}
c_1=(\beta_2^2+2\beta_3^2)^{-1/2}\phantom{mmm}
c_2=(\beta_2^2+5\beta_3^2)^{-1/2}
\label{E:3.8}
\end{equation}
With this choice, $G$ has the correct limit when $\beta_3\rightarrow
0$.
Moreover, all non diagonal elements of $\cal G$ involving either
$\beta_2$ or $\beta_3$ and one of the derivative of the other
intrinsic amplitudes or $q_3$ are exactly zero. Other non diagonal
elements are small at least of the first order and it is possible to
show~\cite{biz0} that they have negligible effects, apart from the
elements of the last line and column. The latter, in fact, are also
small of the first order, but must be compared with the diagonal
element ${\cal J}_3$, which is small of the second order.
{\renewcommand{\tabcolsep}{3.9pt}
\begin{table}[t]
\caption{\label{T:1} The  matrix of inertia $\mathcal{ G}$: 
leading terms and relevant first-order terms.
Other first-order terms are indicated with
the symbol $\approx\!0$.
Here $\gamma = \sqrt{5}\gamma_2\!-\!\gamma_3 $,
$\mathcal{ J}_1= 3(\beta_2^2 + 2\beta_3^2) +2\sqrt{3}(\beta_2^2
\gamma_2 + \sqrt{5} \beta_3^2 \gamma_3)$; 
$\mathcal{ J}_2 = 3(\beta_2^2 + 2\beta_3^2) -2\sqrt{3}(\beta_2^2
\gamma_2 + \sqrt{5} \beta_3^2 \gamma_3)$; and
$\mathcal{ J}_3 = 4(\beta_2^2 \gamma_2^2 +\beta_3^2 \gamma_3^2)
+18(X^2 + Y^2) + 2 (\eta^2 + \zeta^2) 
+8 \xi^2$.
}
{\begin{tabular}{l|ccccccccc|ccc}
\hline
& 
$\dot{\beta_2}$ & 
$\dot{\gamma_2}$ & 
$\dot{\beta_3}$ & 
$\dot{\gamma_3}$ & 
$\dot{X}$ & 
$\dot{Y}$ & 
$\dot{\xi} $ & 
$\dot{\eta}$ &  
$\dot{\zeta}$ & 
$q_1$ & 
$q_2$ & 
$q_3$ \\
\hline
$\dot{\beta_2}$ & 1 & 0  & 0  & 0  & 0  & 0  & 
0 & 0 & 0 & $\approx\!0$ & 
$\approx\!0$ & 0\\
$\dot{\gamma_2}$ & 0  & 
$\beta_2^2$
 &0 & 0  & 0  & 0  & 0  & 0  & 0 & 
$\approx\!0$ & 
$\approx\!0$ &
  $ \frac{-\!\sqrt{40}\beta_2 \beta_3 \xi
}{\sqrt{\beta_2^2+5\beta_3^2}}$ \\
$\dot{\beta}_3$ & 0 & 0 & 1 &
0  & 0 & 0  & 0 & 0 & 0 &
$\approx\!0$ &
$\approx\!0$ &
 0\\
$\dot{\gamma_3}$ & 0  & 0  & 0 &
 $\beta_3^2$  & 
$\approx\!0$   & 
$\approx\!0$  &  0  & 
$\approx\!0$ & 
$\approx\!0$  &  
$\approx\!0$ &
$\approx\!0$ &
$\frac{\sqrt{8}\beta_2 \beta_3 \xi
}{\sqrt{\beta_2^2+5\beta_3^2}}$\\[6pt]
$\dot{X}$ & 0  & 0  & 0  & 
$\approx\!0$ & 
2  & 0  &  0  &
$\approx\!0$  & 0  &  
$\approx\!0$ & 
$\approx\!0$ &  
 $6Y$ \\
$\dot{Y}$ & 0  & 0  & 0  & 
$\approx\!0$ 
 & 0  & 2 &  0  & 0  &
$\approx\!0$  &  
$\approx\!0$ &  
$\approx\!0$ &
$-\!6X$ \\
$\dot{\xi}$ 
& 0 & 0  & 0 & 
 0  & 0  & 0  &  
 2 & 0  & 0  &  
$\approx\!0$  &
$\approx\!0$  &
 $\frac{\sqrt{8}\beta_2 \beta_3 \gamma 
}{\sqrt{\beta_2^2+5\beta_3^2}}\phantom{.}$\\
$\dot{\eta}$ & 
 0 &  0  & 0 &
$\approx\!0$  & 
$\approx\!0$ &
 0  & 0 & 2 & 0  & 
$\approx\!0$  & 
$\approx 0$ &
 $2 \zeta $\phantom{.}\\[6pt]
$\dot{\zeta}$ & 
0 &  0  &  0 &
$\approx\!0$  & 0  &
$\approx\!0$  &  0  & 0  &
 2 &
$\approx\!0 $ &
$\approx\!0$ & 
 $-2 \eta $ \\[6pt]
\hline
$q_1$ & $\approx\!0$  & $\approx\!0$  & $\approx\!0$  & $\approx\!0$  
& $\approx\!0$  & $\approx\!0$  & $\approx\!0$  & $\approx\!0$  
& $\approx\!0$ &
 ${\mathcal{ J}_1}$  & 0  &  0 \\
$q_2$  & $\approx\!0$  & $\approx\!0$  & $\approx\!0$  & $\approx\!0$ 
 & $\approx\!0$  & $\approx\!0$  & $\approx\!0$  & $\approx\!0$  
& $\approx\!0$  
& 0 & 
${\mathcal{ J}_2}$  &
0 \\
$q_3$ & 0  & 
$[..]$ & 0  &
$[..]$ & 
 $6Y$  & 
 $-\!6X$  & 
$[..]$ &  
 $2 \zeta $\phantom{.}&  
 $-2 \eta $  & 
0 & 0 & 
 $\mathcal{ J}_3$ \\
\hline
\end{tabular}
}
\end{table}
}

Relevant terms of the matrix $\cal G$ are shown in the Table~\ref{T:1}.
It can be observed that the variables $\gamma_2$ and $\gamma_3$
appear only in the combination $\gamma = \sqrt{5}\ \gamma_2-\gamma_3$.
In the place of $\gamma_2$ and $\gamma_3$ it will be convenient to use
 the new variable $\gamma$ and the orthogonal
combination $\gamma_0\!=\!c_0\ (\beta_2^2 \gamma_2 +\sqrt{5}\beta_3^2
\gamma_3)$, entering in the expression of $G$. Once more, the factor
$c_0$ is -- at the moment -- an arbitrary function of $\beta_2$ and 
$\beta_3$.
We can observe that $\gamma_0$ is a measure of the triaxiality of the 
overall tensor of inertia. In fact, if $\gamma_0=0$, ${\cal J}_1={\cal
J}_2$ (apart from terms of the second order) and the tensor
of inertia is axially symmetric, even if the nuclear surface is not.
In some way, $\gamma_0$ plays a role similar to that of $\gamma_2$ in
the pure quadrupole case.

Now, to characterize the different degrees of freedom with
respect to the angular momentum projection $L_3$ and to the parity,
it is convenient to perform another change of
representation:
\begin{eqnarray}
X &=& w \sin \vartheta \phantom{mmmj}
Y\ =\ w \cos \vartheta \nonumber \cr
\eta &=&  v \sin \varphi \phantom{mmmm}
\zeta\ =\ v \cos \varphi \\*
\xi &=&   u \sin \chi  \phantom{mmmm}
\gamma\ =\ \sqrt{2}\ \left( \sqrt{\beta_2^2+5\beta_3^2 }\ /
\ \beta_2 \beta_3 \right) u \cos \chi \cr
\frac{\gamma_0}{c_0} &=& f_0(\beta_2,\beta_3) 
\ u_0
\label{E:3.9}
\end{eqnarray}
Choosing $f_0(\beta_2,\beta_3)=\sqrt{\beta_2^2+5\beta_3^2 }$ one
obtains for the determinant of $\mathcal G$
\begin{equation}
G={\rm Det}\;\mathcal{G}=2304\ \big( \beta_2^2+2\beta_3^2 \big) ^2
\ u_0^2\ v^2\ u^2\ w^2\ .
\label{E:3.10}
\end{equation}
At this point, the matrix of inertia takes a very simple form.
The relevant non-diagonal terms are limited to the small sub-matrix
involving $q_3$ and the time derivatives of $\varphi,\ \chi$ and $\vartheta$, 
which can be diagonalized easily. To this purpose,
we evaluate the intrinsic angular momentum component
$L_3=\partial T / \partial q_3$ and also the conjugate
moments $p_\varphi,\ p_\chi,\ p_\vartheta$ of the angular variables 
$\varphi,\ \chi,\ \vartheta$:
\vspace{-2mm}

\noindent
\begin{eqnarray}
p_\varphi &=& 2  v^2\ ( \dot{\varphi} 
+ q_3 ) \nonumber \\
p_\chi &=& 2 u^2\ \left( \dot{\chi} + 2 q_3 \right)  \\
p_\vartheta &=& 2 w^2\ ( \dot{\vartheta} + 3 q_3 ) \ .\nonumber
\label{E:3.11}
\end{eqnarray}
\vspace{-8.5mm}

\noindent
\begin{eqnarray}
L_3 &=& \mathcal{J}_3 q_3 +
\left[  2  v ^2 \dot{\varphi}
+4 u^2 \dot{\chi} 
+6 w^2 \dot{\vartheta} 
\right] \nonumber \\*
&=&(p_\varphi+2 p_\chi+3 p_\vartheta )\ +\ 4 u_0^2 q_3\ .
\label{E:3.12}
\end{eqnarray}
One can immediately observe that when $u_0=0$ ({\it i.e.} when 
${\cal J}_1={\cal J}_2$), $q_3\rightarrow \infty$ unless 
$L_3=\Omega\equiv p_\varphi+2 p_\chi+3 p_\vartheta$. This means that
$\Omega$ is the component of the {\em intrinsic angular momentum}
along the axis 3, 
which survives also if this is an axial--symmetry axis for the 
overall tensor of inertia. 
If we assume
that the potential energy does not depend on $\varphi,\ \chi,$ or
$\vartheta$, it is easy to realize that  $p_\varphi,\ p_\chi$ and
$p_\vartheta$ are quantized and are integer multiples of $\hbar$:
the degrees of freedom corresponding to the pairs of variables
$(v,\ \varphi),\ (u,\ \chi),\ (w,\ \vartheta)$ correspond to 1,
2 and 3 units of angular momentum along the intrinsic axis 3.
This is obvious in the latter case, which concerns the octupole
variables $a^{(3)}_{\pm 3}$, but not for the other two, in which
quadrupole and octupole variables are mixed together.
It is also possible to show~\cite{biz0} that these variables carry 
negative parity.

It remains to consider the variable $u_0$. If we assume that the
differential equation for $u_0$ can be decoupled from the others,
this equation takes the form
\begin{equation}
\left\{ \frac{1}{u_0} \frac{\partial}{\partial u_0}
 \left[u_0 \frac{\partial}{\partial u_0} \right] +
\frac{2}{\hbar^2}
\left[ E_{u_0} - U(u_0) \right]
-\ \frac{1}{u_0^2} \left[ 
\frac{K_0}{2}\right] ^2
 \right\} \phi (u_0) =0
\label{E:3.13}
\end{equation}
where we have put $K_0=L_3-\Omega$. The variable $u_0$ can take positive
as well as negative values. The condition of continuity for the
wavefunction $\phi (u_0)$ and its derivative at  $u_0=0$ imposes
that $K_0=2n_{u_0}$, with $n_{u_0}$ integer. We can conclude that the
degree of freedom associated to the variable $u_0$ carries two units of
angular momentum along the intrinsic axis 3, and it is possible to
show that it carries positive parity.
The inverse of the matrix $\cal G$ turns out to be diagonal (at the
relevant order)
in the space of momenta conjugate to the variables defined in the 
Eq.~\ref{E:3.9} and of the angular momentum components $L_1$,
$L_2$ and $K_0$. The Table~\ref{T:2} shows a more general form of
this matrix, with the variable
$u_0$ replaced by $\gamma_0=u_0/R(\beta_2,\beta_3)$, where 
$R=1/[c_0(\beta_2,\beta_3) f_0(\beta_2,\beta_3)]$
\ (see Eq.~\ref{E:3.9}). A more formal
derivation of these results, involving the derivatives of the Euler
angles, can be found in the Appendix C of ref.~\cite{biz0}.
\begin{center}
\begin{table}[t]
\centering
\caption{\label{T:2} The  matrix ${\cal G}^{-1}$ in the space of momenta
(only the leading terms are shown).
Here, $K_0=L_3-\Omega$,\ with 
$\Omega = p_\varphi + 2 p_\chi + 3 p_\vartheta$, 
and $\gamma_0=u_0/R$.
}
{\footnotesize
{\setlength{\tabcolsep}{2pt}
\begin{tabular}{||l|ccccccccc|ccc||}
\hline
& $\partial /\partial {\beta}$ & $\partial /\partial {\delta}$ &
$\partial /\partial {\gamma_0}$ & 
$\partial /\partial {v}$ & $\partial /\partial {u}$ & $\partial /
\partial {w}$ 
& $\partial /\partial {\varphi} $ & $\partial /\partial {\chi}$ 
&  $\partial /\partial {\vartheta}$ & $L_1$ & $L_2$
& $K_0$ \cr
\hline
$\partial /\partial {\beta}$ & 1 & 0 & 0 & 0 & 0 & 0 & 0 & 0 & 0 & 0 &

0 & 0 \cr
$\partial /\partial {\delta}$ & 0 & $\frac{1}{\beta^2}$ & 0 & 0 & 0 & 
0 & 0 & 0 & 0 & 0 & 0 & 0 \cr
$\partial /\partial {\gamma_0}$     & 0 & 0 & $\frac{1}{R^2}$ & 0 & 0
& 0 & 0 & 0 
& 0 & 0 & 0 & 0 \cr
$\partial /\partial {v}$    & 0 & 0 & 0 & $\frac{1}{2}$ & 0 & 0 & 0 & 
0 & 0 & 0 & 0 & 0 \cr
$\partial /\partial {u}$   & 0 & 0 & 0 & 0 & $\frac{1}{2}$ & 0 & 0 & 
0 & 0 & 0 & 0 & 0 \cr
$\partial /\partial {w}$       & 0 & 0 & 0 & 0 & 0 & $\frac{1}{2}$ & 
0 & 0 & 0 & 0 & 0 & 0 \cr
$\partial /\partial {\varphi}$ & 0 & 0 & 0 & 0 & 0 & 0 & 
$\frac{1}{2v ^2}$  & 0 & 0 &
                                    0   & 0 & 0 \cr
$\partial /\partial {\chi}$    & 0 & 0 & 0 & 0 & 0 & 0 & 0 & 
$\frac{1}{2u^2}$ & 0 & 0  & 0 & 0 \cr
$\partial /\partial{\vartheta}$
                & 0 & 0 & 0 & 0 & 0 & 0 & 0 & 0 & $\frac{1}{2 w^2}$
                                   & 0 & 0 & 0 \\[1mm]
\hline
$L_1$ & 0 & 0 & 0 & 0 & 0 & 0 & 0 & 0 & &
                    $\frac{1}{{\cal J}_1}$ & 0 & 0 \cr
$L_2$ & 0 & 0 & 0 & 0 & 0 & 0 & 0 & 0 & 0 & 0 & 
$\frac{1}{{\cal J}_2}$ & 0 \cr
$K_0$ & 0 & 0 & 0 & 0 & 0 & 0 & 0 & 0 & 0 & 0 & 0 & 
$\frac{1}{R^2\gamma_0^2}$\\[1.5mm]

\hline
\end{tabular}
}}
\end{table}
\end{center}

\section[]{The axial octupole mode with stable quadrupole deformation}
\label{S:4}

This is the simplest case in which the properties of octupole
excitation can be followed from the limit of harmonic oscillations
around the reflection symmetric core to the opposite limit of stable
octupole deformation. A detailed discussion of this subject can be
found in the ref.~\cite{biz0}, and a few preliminary results 
have been reported at two earlier Conferences~\cite{biz1,biz2}.
Here, we only summarize these results.

\begin{figure}[t]
\centerline{\epsfig{file=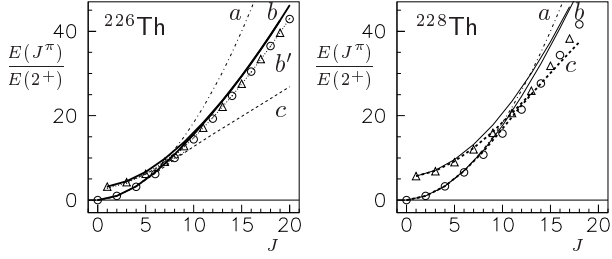,width=80mm,clip=,
bbllx=1,bblly=1,bburx=480,bbury=203}}
\caption{(From ref.~\cite{biz0}). Excitation energies for states of positive
parity (circles) and negative parity (triangles), in units of
$E(2^+)$, for the ground-state band of $^{226}$Th and $^{228}$Th.
Theoretical curves: $a$ -- rigid rotor; $b\ (b^\prime)$ -- present model
with critical
potential, fitted on the $1^-$ state (on the high-spin states); $c$ --
present model with harmonic potential. \label{F:3.1}}
\end{figure}

A preliminary comment is in order. The properties of the quadrupole
vibrations around an axially deformed core are better 
described~\cite{gr1} 
with respect to the intrinsic parameters $a^{(2)}_2$, $\delta a^{(2)}_0=
a^{(2)}_0-\bar{a}^{(2)}_0$ than in terms of the Bohr parameters $\beta_2$ 
and $\gamma_2$.
We have seen that the parameter $u_0$ defined in the eq.~\ref{E:3.9} plays,
in our treatment, a role analogous to that of $a^{(2)}_2$ in the pure
quadrupole Hamiltonian. It appears reasonable, therefore, to use it as
a dynamical variable instead of defining an angle variable similar to
the $\gamma_2$ of the quadrupole case.
Therefore, we can use the form  of the matrix ${\cal G}^{-1}$ given  
in Table~\ref{T:2}, with $R=1$, to derive the differential equation for
$\beta_3$, according to the  Pauli prescriptions (in doing this, we
assume decoupling of the $\beta_3$ motion from the small--amplitude 
oscillations in all other degrees of freedom). One obtains
\begin{equation}
\frac{{\rm d}^2\psi (x)}{{\rm d}x^2} +
\frac{2x}{1+x^2}\frac{{\rm d}\psi (x)}{{\rm d}x}
 + \left[ \epsilon -
\frac{J(J+1)}{6(1+x^2)} -v(x)\right] \psi (x)\!=\!0
\label{E:5.4}
\end{equation}
where $x=\sqrt{2} \beta_3/\bar{\beta}_2$, while $ v(x),\ \epsilon$ are the
potential energy and the energy eigenvalue in a proper energy unit,
and $\psi (-x)=(-)^J \psi (x)$.
As for the potential $v(x)$, we have considered two simple cases: a
quadratic expression $v=\frac{1}{2}cx^2$ or a critical (square-well) 
potential,
as in the X(5) model: $v(x)=0$ for $|x|<b$ and $=+\infty $ for $|x|>b$.
In both cases, the model has one free parameter ($c$ or $b$) to be 
adjusted to fit the experimental data.
In the figure~\ref{F:3.1} the energies of positive and negative parity
levels of $^{226}$Th and $^{228}$Th are compared with different model
predictions. The former turns out to be close to the results we obtain
for a critical-point potential, while the latter is closer to those
obtained with a quadratic potential. 

\section[]{Latest developments}
\label{S:5}
The results reported until now are, apart from some limited extensions,
those contained in our recent paper in Physical Review C~\cite{biz0}. 
From now on, we move to the region where not only $\beta_3$ but also
$\beta_2$ is allowed to vary, in principle down to zero: this is still
a partially unexplored land, were work is 
in progress and results might be subject to change.
Eventually, it will be useful to move from the Cartesian
representation in terms of $\beta_2,\ \beta_3$ to a polar one, in
terms of the new variables $\beta$, $\delta$ with 
$\beta_2= \beta \cos \delta$, $\beta_3= \beta \sin \delta$ as in the
work by Minkov~\cite{minkov}, but for the moment we will continue to use
 the Cartesian representation.

For consistency with the well established results on
the pure quadrupole case, we need that our results converge to those of
the Bohr model in the limit of small octupole deformation.
As we are going to see, this requirement will put some restrictions 
on the possible models of octupole plus quadrupole excitations.
{\it E.g.}, the form of the matrix $\cal G$ we have used in the case
of permanent quadrupole deformation is not suitable in the present
case, as its determinant $G$, at the limit of small $\beta_3$, is
proportional to $\beta_2^4$, while it should be proportional to 
$\beta_2^8$ according to the Bohr model.
It is easy to recognize that the responsibility for the disagreement
can be attributed to the choice of $u_0$ as a dynamical variable.
Since in the Bohr model the variable $\gamma_2$ is used in the place of
$a^{(2)}_2$, it is now necessary to replace $u_0$ with an adimensional
variable $\gamma_0=u_0/R$. We could choose now  
$R=\beta\equiv \sqrt{\beta_2^2+\beta_3^2}$, to obtain the correct
limit $G\propto \beta_2^8$ for $\beta_3\! \rightarrow 0$.
This is somewhat better, but still not enough:
the equation for $\beta_3$ obtained with the Pauli 
quantization rule does not converge to the one of Bohr when
 $\beta_3\rightarrow 0$.

In fact, if we only take into account the dynamical variables
$\beta_2$ and $\beta_3$ (assuming that their equations can be
approximately separated from that of $\gamma_0$, as in Iachello
X(5) model, and from all other dynamical variables), we obtain
\begin{eqnarray}
\left\{ G^{-1/2}\frac{\partial}{\partial \beta_2} \left[ G^{1/2} 
\frac{\partial}{\partial \beta_2 }\right] \right. 
+G^{-1/2}\frac{\partial}{\partial \beta_3} \left[ G^{1/2} 
\frac{\partial}{\partial \beta_3 }\right]&& \cr
+\left. \frac{2}{\hbar^2} \left[ E - V_0(\beta_2,\beta_3) \right] - 
\frac{J(J+1)}{3(\beta_2^2+2\beta_3^2)} \right\}\ \Psi (\beta_2,
\beta_3) &=&0
\end{eqnarray}
This equation can be somewhat simplified with the substitution
\begin{equation}
\Psi (\beta_2,\beta_3) = g^{-1/2}\ \Phi (\beta_2,\beta_3)
\label{E:5.11}
\end{equation}
with $g\propto G^{1/2}$, to obtain
\begin{equation}
\bigg\{
\frac{\partial^2}{\partial \beta_2^2} 
+ \frac{\partial^2}{\partial \beta_3^2}
+ \frac{2}{\hbar^2} \left[ E - V_0 \right] 
-\frac{J(J+1)}{3(\beta_2^2+2\beta_3^2)} 
\ +\ v(\beta_2,\beta_3) \bigg\}\ \Phi (\beta_2,\beta_3) =0
\label{E:5.12}
\end{equation}
with
\begin{equation}
v(\beta_2,\beta_3) =\frac{1}{4g^2}\  \left[ \left( \frac{\partial g
}{\partial \beta_2} \right) ^2 +
\left( \frac{\partial g}{\partial \beta_3} \right) ^2 \right]
- \frac{1}{2g} \left[ \frac{\partial^2 g}{\partial \beta_2^2}
+ \frac{\partial^2 g}{\partial \beta_3^2} \right]
\label{E:5.13}
\end{equation}
If $R=\beta$,\ $g=(\beta_2^2+\beta_3^2)\ (\beta_2^2+2\beta_3^2)$,
and one obtains
\begin{equation}
v(\beta_2,\beta_3)=-\frac{5\beta_2^4+16 \beta_2^2\beta_3^2
+14\beta_3^4}{(\beta_2^2+ \beta_3^2)\ (\beta_2^2+ \beta_3^2)^2}
\Rightarrow -\frac{5}{\beta_2^2}\ \textit{\rm for}\ \beta_3 \Rightarrow
0
\label{E:5.14}
\end{equation}
A similar calculation is possible for the pure quadrupole case,
 starting from the Bohr expression of $G\propto 
\beta_2^8$, but the result is $v(\beta_2)= -2/\beta^2$.
Even if the limit of $G$ for $\beta_3\rightarrow 0$ converge to the
corresponding one of the Bohr model, this is not necessarily true for
$v$.

However, it is easy to realize that $v(\beta_2,\beta_3)$
converges to $v(\beta_2)$ of the Bohr model if the first and second
partial derivatives of $g$ with respect to $\beta_3$ tend to zero
when  $\beta_3\rightarrow 0$.
A possible choice of variables leading to this result corresponds to
keeping $c_0=\sqrt{\beta_2^2+2\beta_3^2}$ (instead of 
$\sqrt{\beta_2^2+5\beta_3^2}$\ ) in the definition of $\gamma_0$.
In this case one obtains
\begin{equation}
g(\beta_2,\beta_3)= \frac{(\beta_2^2 + \beta_3^2)
 (\beta_2^2 + 2 \beta_3^2)^2}{\beta_2^2 + 5 \beta_3^2} \Rightarrow
\beta_2^4 \left[ 1+8 \left( \beta_3/\ \beta_2\right) ^4 + ... \right]
\end{equation}
and the first and second derivative of $g$ with respect to $\beta_3$ 
vanish for $\beta_3\rightarrow 0$.

\section[]{Specific model for quadrupole--octupole oscillations}
\label{S:6}

We now consider the case of simultaneous quadrupole--octupole
oscillations, and in particular the quadrupole motion corresponding to
the critical point of phase transition described by the X(5) model.
At the limit of small amplitude for the octupole oscillations, we
should therefore obtain the same results of X(5). The point is that
the octupole amplitude must be, at this limit, small {\em compared to
the quadrupole amplitude}, which, in turn, can become zero. It is
convenient, therefore, to use the new variables $\beta$, $\delta$
defined by
\begin{equation}
\beta_2=\beta \cos{\delta}\phantom{mmmm}\beta_3=\beta \sin{\delta}
\label{E:6.1}
\end{equation}
and reach the limit by confining $\delta$ to very small values by a 
proper potential term.
It is clear that also the amplitudes $v,\ u,\ w$ must be 
small compared to $\beta$, as well as $\gamma_0$ compared to 1.
At the moment, however, we will forget their
presence and only discuss the Schr\"odinger equation involving 
the variables $\beta,\ \delta$ and the Euler angles. With this 
\textit{ansatz}, and assuming that the potential energy has the form
$(\hbar^2/2)\ V(\beta, \delta )$,  for $K=0$ we obtain
\begin{equation}\Bigg\{ \frac{1}{g}\frac{\partial}{\partial \beta} 
\left[ g
\frac{\partial}{\partial \beta} \right]
+\frac{1}{g}
\frac{\partial}{\partial \delta} \left[ \frac{1}{\beta^2} g
 \frac{\partial}{\partial \delta} \right]
+ \epsilon - V
- \frac{A_J}{\beta^2 (1+\sin^2\delta)}
\ \Bigg\}\ \Psi (\beta ,\delta )=0
\label{E:6.4}
\end{equation}
where  $g\propto G^{1/2}\propto \beta^5\ [(1 + \sin ^2\delta )^2/\ (1 + 
4\sin ^2\delta )]\ \gamma_0\ u_0\ v\  u\ w$,
$\epsilon = 2E/\hbar^2$,  and
$A_J=J(J+1)/3$.

The Eq.~\ref{E:6.4} has a structure very similar to that of the Bohr
equation for pure quadrupole motion at the limit close to the axial
symmetry, with our parameter $\delta$ in the place of $\gamma_2$.
In the case of X(5) symmetry, the Bohr equation has been 
solved~\cite{iac2} by approximate separation of the variables,
substituting the factor $1/\beta_2^2$ with a proper 
average value in the differential
equation for $\gamma_2$. It has already been noted, however, that this
approach does work for the $\gamma$ excited bands but not 
for the negative-parity ones~\cite{biz104}. 
Now we want to explore 
some alternative procedure which could 
better account for  the experimental data.
 
\begin{figure}[t]
\centerline{\epsfig{file=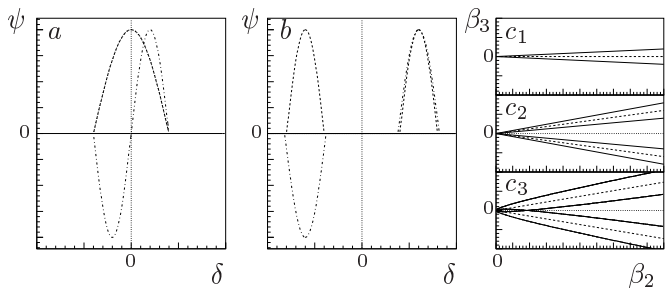,width=89mm,
clip=,bbllx=30,bblly=497,bburx=520,bbury=704}}
\caption{The lowest even and odd parity wavefunctions
 in a separable model with $\delta$ restricted to a small
interval around 0 (part $a$) or around $\pm \delta_0$
(part $b$). The part $c$ shows the corresponding allowed
regions in the $\beta_2 - \beta_3$ plane ($c_1$ and $c_2$,
respectively) and the region corresponding to one standard
deviation for the localized wavefunctions in the schematic
model discussed in the text ($c_3$). \label{F:3.3}}
\end{figure}

It is convenient to exploit the result of
Eq.s~\ref{E:5.11},\ref{E:5.12},\ref{E:5.13}, to eliminate in the 
Eq.\ref{E:6.4} the first--derivative terms, with the substitution
\begin{equation}
\Psi(\beta ,\delta )= g^{-1/2}\ \Psi_0 (\beta ,\delta)
\label{E:6.5}
\end{equation}
giving
\begin{eqnarray}
&\Bigg\{ & \frac{\partial^2}{\partial \beta^2} + \frac{1}{\beta^2}  
\frac{\partial^2}{\partial \delta ^2} 
+ \epsilon - V(\beta ,\delta) - \frac{A_J}{\beta^2 (1+\sin^2\delta)}\\
&&-\frac{7}{4}\ \frac{1}{\beta^2} 
-\frac{2\sin ^2\delta\ (25+12\sin ^2\delta +8\sin ^4\delta)}{
(1+\sin ^2\delta)(1+4\sin ^2\delta)^2} \ \frac{1}{\beta^2}\Bigg\} \
\Psi_0 =0\ .\nonumber
\label{E:6.6}
\end{eqnarray}
We can separate the potential term in two parts, one of which depends
only on $\beta$: $V(\beta, \delta )=V_0(\beta ) +V_1(\delta ,\beta)$
 and search a solution in the form 
$\Psi_0(\beta,\delta )=\psi(\beta )\Phi(\delta, \beta)$:
\begin{eqnarray}\label{E:6.7}
\frac{1}{\psi}&&\hspace{-6mm}\left[ 
\frac{\rm{d}^2 \psi}{\rm{d} \beta^2} +
\left( \epsilon - V_0(\beta ) - 
\frac{A_J+\frac{7}{4}+A^\prime}{\beta^2} \right)
\psi \right] \\*
&+&\frac{1}{\beta^2 \Phi} \left[ \frac{\partial^2 \Phi}{\partial
\delta^2}
+\frac{2\beta^2}{\psi} \frac{\partial \psi}{\partial \beta} 
\frac{\partial \Phi}{\partial \beta}
+\beta^2 \frac{\partial^2 \Phi}{\partial \beta^2}
+ \Bigg( A^\prime - \beta^2\ V_1(\delta ,\beta) \right. \nonumber \\*
&+&\left. \frac{A_J \sin ^2\delta}{1+\sin ^2\delta} 
-\frac{2\sin ^2\delta\ (25+12\sin ^2\delta +8\sin ^4\delta)}{
(1+\sin ^2\delta)(1+4\sin ^2\delta)^2} 
\Bigg) \Phi \right] =0\ .\nonumber
\end{eqnarray}
This equation is exactly separable if $\beta^2 V_1(\delta ,\beta)$
does not depend on $\beta$, {\it i.e.} 
$V_1(\delta ,\beta) = v(\gamma)/\ \beta^2$. This class of potentials 
has been considered, {\it e.g.}, by Fortunato~\cite{fortunato} in his 
general discussion about the solutions of the Bohr Hamiltonian.
We can reasonably suspect that it be not realistic in our
case, but a discussion of this simply solvable model can shed light
on the general properties of the quadrupole--octupole vibrations and
help to identify more realistic solutions.
With such a potential, $\Phi= \phi(\delta)$ is a solution of the 
differential equation
\begin{eqnarray}
\left[ \frac{{\rm d}^2\!\phi}{{\rm d}\delta^2}
\!+\!\left(\!A^\prime\!-\!v(\delta)\!+
\!\frac{A_J \sin ^2\!\delta}{1\!+\!\sin ^2\!\delta} 
\!-\!\frac{2\!\sin ^2\!\delta (25\!+\!12\sin ^2\!\delta \!+
\!8\sin ^4\!\delta)}{
(1\!+\!\sin ^2\!\delta)(1\!+\!4\sin ^2\delta)^2} 
\right)\!\phi \right]\!=\!0
\label{E:6.8}
\end{eqnarray}
We want to explore first the case of small oscillations of $\delta$
around zero. For our discussion, we do not need to specify the exact
form of the potential, but, just as an example, we can assume
a harmonic restoring force and expand up to the second order the terms
in $\sin^2\delta$ (whose effect, however, will be negligible at least 
for not-to-high values of $J$), to 
obtain the eigenvalues
$A^\prime_n=(2n+1) A^\prime_0$
and eigenfunctions alternatively even or odd, with parity
$(-)^n$.
Or, as an alternative, one could use a square-well potential
$v(\delta)=0$ for $\delta< \delta_\ell $ and $=+\infty$ elsewhere,
as exemplified in Fig.~\ref{F:3.3}~$a$.
Other potentials would give a different spectrum of eigenvalues
but, if they have a single minimum at $\delta=0$, the ground state must
correspond to a symmetric solution. We indicate with $A^\prime_+$
the eigenvalue corresponding to the ground state and with $A^\prime_-$
the one corresponding to the lowest antisymmetric solution.
We want that, for even spin and parity, the equation in $\beta$ 
take the form of that of the X(5) model,
{\it i.e.}, in our notations,
\begin{equation} 
\frac{\rm{d}^2 \psi}{\rm{d} \beta^2} +
\left( \epsilon - V_0(\beta ) - 
\frac{A_J+2}{\beta^2} \right)
\psi  =0
\label{E:6.10}
\end{equation}
Comparison with Eq.~\ref{E:6.7} shows that this result would be
obtained with $ A^\prime_0=1/4$. Actually, it is probably unnecessary
to assume that this relation holds. A general problem of all models
 considering explicitly only
part of the overall set of dynamical variables, is the effect of the
zero--point energies of the neglected degrees of freedom, which
possibly depend on the value of the active model variables. 
If a phenomenological
potential is used for the latter, this potential should already
include the zero--point energies of all other degrees of freedom not
explicitly taken into account.

Assuming that the positive--parity states must be described by the
X(5) Hamiltonian, the following equation in $\beta$ 
holds for the lowest $K^\pi=0^\pm$ bands
\begin{equation} 
\frac{\rm{d}^2 \psi}{\rm{d} \beta^2} +
\left( \epsilon - V_0(\beta ) - 
\frac{A_J+2+\Delta_\pi}{\beta^2} \right)
\psi  =0
\label{E:6.11}
\end{equation}
We now consider in particular the case of the critical--point potential,
$V_0=0$ when $\beta$ is in the interval $0- \beta_0$ and $V_0=
+\infty$ elsewhere. As for the parity dependent term $\Delta_\pi$,
with our assumptions is  $\Delta_+=0$. For negative parity, 
the term $\Delta_- = A^\prime_- - A^\prime_+$ 
will be considered as an adjustable parameter.
With this assumption, the spectrum of eigenvalues is given by
\begin{equation}
\epsilon(s,J,\pi)-\epsilon_0= C\ [x(s,J,\pi)]^2
\label{E:6.14}
\end{equation}
with $C$ constant, $x(s,J,\pi)$ the $s^{th}$ zero of the Bessel function
$J_\nu(x)$ and
\begin{equation}
\nu = \sqrt{J(J+1)/3+9/4+\Delta_\pi}
\label{E:6.15}
\end{equation}
\begin{figure}[t]
\centerline{\epsfig{file=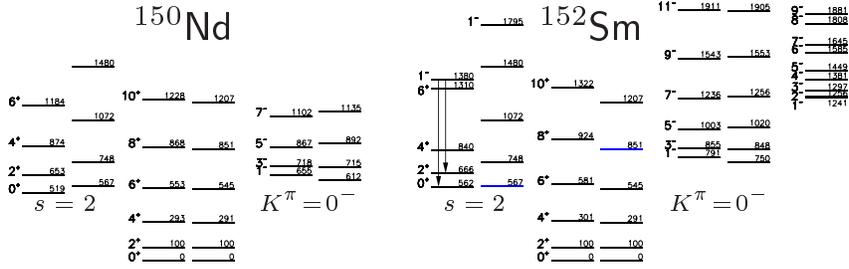,width=112mm,clip=,
bbllx=29,bblly=555,bburx=540,bbury=714}}
\caption{The experimental (left) and theoretical energies 
(normalized to that of the first excited state) of the $K=0$
bands in $^{150}$Nd and  $^{152}$Sm. In the latter, the experimental
values for the $K^\pi=1^-$ band are also shown. Theoretical values
for even $J$ and parity are those of the X(5) model, for odd $J$
and parity are obtained with $\Delta_-=15$ for  $^{150}$Nd and
with  $\Delta_-=20$ for $^{152}$Sm. \label{F:3.2}}
\end{figure}
Fig.~\ref{F:3.2} shows a partial level scheme of $^{152}$Sm and
$^{150}$Nd, normalized to the energy of the first excited state, and
compared with the values derived from Eq.~\ref{E:6.14} (which,
for positive--parity states, coincide with those of the X(5) model).
The agreament is fairly good in both cases. In $^{152}$Sm, the
comparison can be extended to the lowest negative--parity state of 
the $s=2$ band, if the level reported as $1^{(-)}$ really belongs to 
this band as it would be suggested by its decay. If it is  so, its
energy is significantly lower than the model prediction, but this
happens also for all the $s=2$ excited states of even parity and spin.
It must also be noted that 
in $^{152}$Sm the odd-$J$ levels of the next octupole band
($K^\pi=1^-$) seem to mix with those of the lowest one, as indicated
by the large odd-even staggering. In the heavier $N=90$ isotones the
second octupole band (with $K\neq 0$) gets closer to the first one, 
and our model
is no longer valid in such a case. In well established X(5) nuclei 
lying in different regions, the lower negative bands seem to be built
over intruder states, having a band head $J^\pi=3^-$ or $5^-$.

Is the agreement shown by Fig.~\ref{F:3.2} a purely accidental
one? It is possible, but at least
 it shows that {\em in this case} experimental data can be
reproduced by using a separable potential. We shall see that this is
{\em not} possible in other cases, and in particular in those which seem
more interesting to us, $^{224}$Th and $^{224}$Ra.

In these two nuclei, the even $J$, positive parity states follow
closely the X(5) model predictions. The negative--parity part of
the ground state band lies significantly higher than the positive
parity part at low values of $J$, while the two parts merge together
above $J \approx 8$. This behaviour is very different from those of 
the $\gamma$--excited bands in the X(5) nuclei and of the negative
parity bands in $^{152}$Sm and $^{150}$Nd.
It could be natural to assume that in these mass-224 nuclei the 
axial quadrupole and octupole amplitudes do not oscillate
independently, but are strictly correlated 
together\footnote{Such a behaviour has been predicted as possible 
in this region by Nazarewicz and Olanders calculations
in the frame of the Strutinski model~\cite{nazar}.}: the nuclear
deformation oscillates along two symmetric valleys, starting at 
$\beta=0$ and caracterized by an average value of $\delta$
close to $+ \delta_0$ or to $-  \delta_0$. With reference to our
Eq.s~\ref{E:6.1}, this means that $\beta$ oscillates in a wide 
interval, while $\delta$ is confined to a narrow region around some
{\em nonzero value}. As the potential must be symmetric with respect
to $\delta$, this means that possible values of $\delta$ are localized
in one of the {\em two} regions around $+\delta_0$ or $-\delta_0$.
According to Eq.~\ref{E:6.8} (with $v(\delta)$ indepentent of
$\beta$), this would result in a negligible parity staggering.

In fact, as long as we forget the symmetry of the wavefunction with 
respect to $\delta$, we find a doubly degenerate spectrum,
corresponding to eigenfunction localised either around $+\delta_0$
or around $-\delta_0$. Reflection symmetry requires that the complete
wavefunction be symmetric (the sum of the two) for even values of J and 
$\pi$, 
or antisymmetric (their difference) for odd values. As long as the two
localized wavefunctions have no overlap (or have a negligible one),
the eigenvalues corresponding to even or odd combinations are (almost)
equal to those of the localized solutions. The  dependence on $J$
of the last-but-one term of Eq.~\ref{E:6.8} has a minor effect
on the results, particularly  at low values of $J$.
As a consequence, the ground--state band has no staggering
(or a very limited one).

How can we proceed? A possible way out is to assume a potential,
symmetric in $\delta$, and approaching the harmonic behaviour around 
$\delta \approx \pm \delta_0$. In this case, the equation is no
longer separable in the entire field. For large values of $\beta$,
one still finds localized solutions, with a $\beta$ dependent
extension around the average value ($\pm \delta_0$): 
for a fixed value of $\beta$, the variance of $\delta$ would
have the form $D^2(\delta)= C/\beta$. As $\beta \rightarrow 0$, the
two localized solutions start to overlap (Fig.~\ref{F:3.3}~$c_3$), 
and this would remove the
degeneracy of even and odd combinations, as long as the wavefunctions
in $\beta$ have a non-negligible value in the region where the overlap
is significant. At moderately large values of $J$, the
significant
values of $\psi(\beta )$ are pushed out of this region by the
centrifugal-like term of Eq.~\ref{E:6.11}, and the staggering will
disappear.

Although this approximation is certainly insufficient just
in the region of overlap, we think it can give a qualitative
explanation of the behaviour of the negative--parity band
(not far from the point of view of Jolos and von 
Brentano~\cite{jolos}) and, perhaps, give some indication for
future developments.


\begin{thebibliography}{99}

\bibitem{biz0} P.G. Bizzeti, and A.M. Bizzeti-Sona (2004) {\em
Phys. Rev. C} {\bf 70} 064319.

\bibitem{biz146} L. Bargioni {\it et al.} (1996) {\em
Phys. Rev. C} {\bf 51} R1057.

\bibitem{biz148} Zs. Podoly\'ak {\it et al.} (2000) {\em
Europ.Phys. J. A} {\bf 8} 147.

\bibitem{iac1} F. Iachello (2000) {\em Phys. Rev. Lett.} {\bf 85} 3580.

\bibitem{iac2} F. Iachello (2001) {\em Phys. Rev. Lett.} {\bf 87} 052502.

\bibitem{iac3} F. Iachello (2003) {\em Phys. Rev. Lett.} {\bf 91} 132502.


\bibitem{z0} D. Bonatsos {\it et al.} (2004) {\em Phys. Lett. B} 
{\bf 588}  172 

\bibitem{casten} R.F. Casten and N.V. Zamfir, {\em Phys. Rev. Lett.} {\bf 87} 052503.

\bibitem{biz104} P.G. Bizzeti, and A.M. Bizzeti-Sona (2002) {\em
Phys. Rev. C} {\bf 66} 031301.

\bibitem{biz1} P.G. Bizzeti, and A.M. Bizzeti-Sona (2004) {\em
Eur. Phys. J. A} {\bf 20} 179.

\bibitem{biz2} P.G. Bizzeti (2003) in {\em Symmetries in Nuclear
Structure} (ed. A. Vitturi and R. Casten; World Scientific, Singapore)
p. 262.
 
\bibitem{hutter} C.Hutter {\it et al} (2003) {\em Phys. Rev. C}
{\bf 67} 054315.

\bibitem{ivbm} A.I. Georgieva, H.G. Ganev, J.P.~Draayer (2005);
 H.G. Ganev, A.I. Georgieva, J.P.~Draayer (2005);
V.P. Garistov, A.A. Solnyshkin,  A.I. Georgieva, 
V.V. Burov,  H.G. Ganev (2005) {\em These Proceedings}.

\bibitem{spdf} C.E. Alonso {\it et al.} (1995) {\em Nucl. Phys. A}
{\bf 586} 100; A.A. Raduta and D. Ionescu (2003) {\em Phys. Rev. C}
{\bf 67} 044312; N.V. Zamfir and D. Kunezov (2003) {\em Phys. Rev. C}
{\bf 67} 014305; and references therein.

\bibitem{donner} W. Donner and W. Greiner (1966) {\em Z. Phys.} {\bf
197}  440.

\bibitem{models} P.A. Butler and Nazarewicz (1996)  {\em Rev. Mod. Phys.}
{\bf 68} 349, and references therein.
 
\bibitem{jolos} R.V. Jolos and P. von Brentano (1999) {\em Phys. Rev. C}
{\bf 60} 064317.

\bibitem{modelsr} N. Minkov {\it et al.} (2001) {\em Phys. Rev. C}
{\bf 63} 044305;  N. Minkov {\it et al.} (2004) {\em Proc. of the 23th
Int. Workshop on Nuclear Theory (ed. S. Dimitrova, Sofia)} p. 203.

\bibitem{wexler} C. Wexler and G.G. Dussel (1999)  {\em Phys. Rev. C}
{\bf 60} 014305.

\bibitem{pauli}  W. Pauli (1933) in {\em Handbuch der Physik}
(Springer, Berlin) Vol. XXIV/I.

\bibitem{minkov} N. Minkov, {\em These Proceedings}.

\bibitem{eisen1} J. Eisenberg and W. Greiner (1987) {\em Nuclear Models}
(North Holland, Amsterdam).

\bibitem{gr1} See Section 6.1 of ref.~\cite{eisen1}.

\bibitem{fortunato} L.Fortunato (2005) {\em These Proceedings}.

\bibitem{nazar} W. Nazarewicz and P. Olanders (1985) {\em
Nucl. Phys. A} {\bf 441} 420.

\end{thebibliography}
\end{document}